\documentclass{article}

\usepackage[preprint,nonatbib]{neurips_2023_gaied}

\usepackage[utf8]{inputenc} 
\usepackage[T1]{fontenc}    
\usepackage{hyperref}       
\usepackage{url}            
\usepackage{booktabs}       
\usepackage{amsfonts}       
\usepackage{nicefrac}       
\usepackage{microtype}      
\usepackage{xcolor}         
\usepackage{enumitem}
\usepackage{graphicx}
\usepackage{caption}
\usepackage{subcaption}
\usepackage[export]{adjustbox}
\usepackage{csquotes}       

\title{Are LLMs Useful in the Poorest Schools?\\TheTeacher.AI in Sierra Leone}

\author{%
  Jun Ho Choi \\
  Columbia University \\
  \texttt{jc5341@columbia.edu} \\
  \And
  Oliver Garrod \\
  FabData \\
  \texttt{oliver.garrod@fabdata.io} \\
  \And
  Paul Atherton \\
  FabInc \\
  \texttt{paul@fabinc.co.uk} \\
  \And
  Andrew Joyce-Gibbons \\
  FabData \\
  \texttt{andrew.joyce-gibbons@fabdata.io} \\
  \And
  Miriam Mason-Sesay \\
  EducAid Sierra Leone\\
  \texttt{miriam@educaid.org.uk} \\
  \And
  Daniel Bj\"{o}rkegren \\
  Columbia University \\
  \texttt{dan@bjorkegren.com} \\
}

\begin{document}

\maketitle

\begin{abstract}
Education systems in developing countries have few resources to serve large, poor populations. How might generative AI integrate into classrooms? This paper introduces an AI chatbot designed to assist teachers in Sierra Leone with professional development to improve their instruction. We describe initial findings from early implementation across 122 schools and 193 teachers, and analyze its use with qualitative observations and by analyzing queries. Teachers use the system for lesson planning, classroom management, and subject matter. Usage is sustained over the school year, and a subset of teachers use the system more regularly. We draw conclusions from these findings about how generative AI systems can be integrated into school systems in low income countries.

\end{abstract}

\section{Introduction}

Developing countries have dramatically increased school attendance in the last decades. However, the quality of instruction is low, and many classrooms rely on rote instruction. Many pupils do not learn the foundations of literacy and numeracy, resulting in 90\% of 10-year-olds being unable to read a simple story \cite{wbpovstate}. Many teachers, particularly in remote areas, face difficulties in accessing support for pedagogical decision-making within the classroom. 

The past year has seen several promising educational applications of Large Language Models (LLMs) for people who are wealthy enough to have digital devices, good internet connectivity, and high baseline levels of literacy and education (such as Khanmigo \cite{khanmigo}). But can LLMs help the world's poorest classrooms?

We focus on Sierra Leone (SL), which (in 2021) had a GDP per capita of \$509.48. By that measure, the average American is 137 times richer \cite{worldeconomicoutlook}. 3.3 million children are enrolled in schools out of a population of 8.4 million - meaning nearly 40\% of the population is currently in school. Teachers are stretched thin, as the average primary school teacher teaches 42 pupils in class, and one in three of them hold no teaching qualifications. Opportunities for professional development are limited given financial constraints: only 40\% of the funding comes from the Government, while the remaining expenses are covered through volunteering or contributions from local communities. Teachers have limited experience with technology and infrastructure is a barrier. In 2020, only 34\% of secondary schools had electricity, and just 8\% had internet access. These numbers are even lower among primary schools (6\% and 1\%). But access to 2G (voice and SMS) is more widespread: 86\% of schools nationally were within range of mobile phone coverage \cite{mullan_edtech_2020}.

Given these challenges, our intervention focused on how we can use the advances in LLMs to help teachers and overcome some of these challenges. This paper describes TheTeacher.AI, a chatbot developed to train and support teachers through conversational interactions. Powered by GPT-3.5 Turbo, it is tailored for local conditions and accessible through WhatsApp, accommodating poor network connectivity and technical experience.

\section{Context}

Since the introduction of Free Quality School Education in 2018, there has been a large increase in the number of children attending school, with enrolment rising from 2m to 3.3m in the last four school years \cite{ASC21}. This surge in enrolment has placed a strain on teachers, as while average class sizes range from 31 to 59 students per teacher, many are still unpaid and unqualified \cite{ASC21}.

Teachers are the backbone of any education system, playing a critical role in facilitating learning. They also represent a substantial proportion of education budget expenditure, accounting for three-quarters of the budget at the primary level in developing countries \cite{worldbank2018learning}. Despite their pivotal role in improving foundational literacy and numeracy, teachers often lack sufficient support, both in terms of professional development and in classroom assistance, such as lesson planning and answering queries. This exacerbates the severity of the learning crisis. Many roads become impassible during the rainy season, making it challenging to provide support in person. Furthermore, traditional training requires taking them out of schools and often traveling long distances to face-to-face workshops, which are expensive and consequently, infrequent.

\section{Implementation}

TheTeacher.AI was developed to support teachers with an AI-powered WhatsApp chatbot that offers on-demand support and training through conversational interactions. The chatbot is a digital helper for teachers, providing information and answering queries. 

Users send WhatsApp text messages to the chatbot, which passes on these queries to OpenAPI's GPT. Before reaching GPT, the said queries are processed by Fab Data to add text on safeguards, best practice pedagogy (guided by the Education Endowment Fund's reports on education evidence \cite{eef}), and context-specific text (see Sections \ref{antijb} and \ref{localisation}, and the system message in Appendix \ref{systemmessage} for details). It was then made available to teachers, free of charge, in an initial pilot phase which utilised existing initiatives. This was an unfunded exercise so was limited in its structured support, but gives an early indication of use. 

\subsection{Constraints and design choices}

\paragraph{Network connectivity is poor and intermittent.}
As of 2022, 93\% of the Sierra Leonean population is covered by 2G mobile signal, but 3G and 4G coverage is more limited to 83\% and 50\%, respectively \cite{gsma2023}. WhatsApp was chosen as the interface for the chatbot, as it performs better under poor network conditions than an app (which would require downloading) or a web interface (which requires a more persistent connection). Additionally, WhatsApp consumes less data, making it cost-effective and accessible. WhatsApp is nearly ubiquitous in Sierra Leone among supported phones. Since many teachers are relatively familiar with WhatsApp, it reduces the barrier to adoption.

\paragraph{Teachers have digital devices; many learners do not.} 
71\% of households own mobile phones. Teachers are likely to own at least a basic phone, and some have access to feature phones or smartphones, but fewer students do. For that reason, and due to safeguarding concerns with new technologies, we chose to focus on teachers.

\paragraph{Inference is costly relative to incomes.}
The cost of using GPT is a significant barrier for individuals with limited financial resources. OpenAI's ChatGPT Plus plan is currently priced at \$20 per month in the U.S., which would account for 52\% of the monthly GDP per capita in Sierra Leone. This can make inference costs unaffordable for many. This is one reason we directed this tool primarily towards educators and limited query topics to teaching and learning.

\paragraph{Local conditions.}
Default GPT responses may better suit U.S. teachers than those in Sierra Leone. Guardrails are necessary to ensure that the responses are both culturally sensitive (e.g. religiously or politically) and pedagogically relevant. Children's safeguarding was considered as paramount. Although corporal punishment was officially banned in September 2021, it remains culturally accepted. Most teachers and parents still view the abandonment of corporal punishment as disastrous to good discipline, and the law is not yet enforced. In our testing, GPT-3.5 was consistent in messaging against corporal punishment, meaning additional safeguards did not appear necessary.

\subsection{Technological setup} \label{techsetup}
Users interact with the system by sending WhatsApp messages to the chatbot's number. These messages are then relayed via Twilio to an API endpoint, which manages the user's chat history, sends a request to OpenAI's Chat completions API, and relays the response back via Twilio.

OpenAI requests take the form of: a system message, followed by the user's chat history (alternating user messages and chatbot responses), followed by the user's latest message, followed by the system message again. We find that repeating the system message for a second time at the end of the request provides far stronger protection against `jailbreaks' without sacrificing fluency of the chat (see \ref{antijb} for more details). The system message is based on a shortened version of Anthropic's `HHH prompt' \cite{askell2021general}, with the examples removed, a list of rules and guidelines related to pedagogical content and the local setting added, and constraints added on what can and cannot be answered. The system message is provided in Appendix~\ref{systemmessage}.

For chat completion we use the model GPT-3.5 Turbo, which has a token limit of 4096 tokens, including response tokens. We prune messages from the start of the chat history whenever this token limit would otherwise be exceeded.%
\footnote{That is, when the combined total number of tokens in the system message (x2) and the user's chat history exceeds 4096 minus a maximum response size (set to 500 tokens).} 

\subsection{Hallucinations and anti-`jailbreak' measures}\label{antijb}
A substantial concern with LLMs is their potential for producing plausible-sounding yet factually incorrect responses, a phenomenon known as hallucination. To address this concern, when designing the system message, two authors collaborated with an external teacher to evaluate the factual accuracy of TheTeacher.AI's responses. They started by creating a set of potential questions that teachers in developing countries might ask to a chatbot, covering topics such as course preparation and discipline management. Each assessor then individually assessed these questions and discussed any problematic ones. The resulting system message incorporates statements aimed at mitigating this issue. However, it remains a concern, especially for populations that may have difficulty validating LLM responses, such as students. Hallucinations should be monitored and guarded against in any future scale-up attempts.

\par Demonstrations of breaking the constraints of ChatGPT's system message appeared almost immediately after its release in November 2022 (e.g. \cite{jailbreak1, jailbreak2}). For example, sending a system message at the start of a chat completion request results in any subsequent user requests intended to override that behavior becoming more recent and typically taking precedence. A surprisingly effective low-cost solution to this problem is simply to send the system message twice -- once at the start of the request and again at the end of the request. We also experimented with sending the system message only once at the end, but this disrupted the flow of the conversation in ways that sending it twice does not. The main drawback to this approach is that the system message takes up double the number of tokens in the request, leaving less space for chat history. However, given our target users, we preferred safety over longer histories.%
\footnote{To verify this method, we generated 10 conversations both with and without system message duplication. In each case, we used the same sequence of 6 attempts to alter the chatbot's behaviour, with a temperature setting of 1.0. The attempts were selected from \cite{jailbreak3}. We then set up a separate chat completion pipeline for the model GPT-4, prompting it to rate each response in each conversation (n=60 for each condition) on a scale of 0 to 10 for its adherence to the system message, where 0 represents no adherence and 10 represents perfect adherence. In the ``no-repeat" condition, the mean rating was 2.4, while in the ``repeat" condition, the mean rating was 9.3.}

\subsection{Tailoring for local conditions}\label{localisation}

We tailored the system so responses were intelligible to teachers, and did not assume access to resources (e.g. digital technology) that may be expected in a U.S. or European classroom.

By default, GPT-3.5 Turbo frequently returns suggestions that require advanced digital technology, such as digital whiteboards or online search engines. We generated 40 responses from default GPT-3.5 Turbo to the query: ``Can you suggest a lesson plan for grade 6 geography. The topic is climate change" and prompted GPT-4 to rate the responses on a scale of 0 to 10 for appropriateness in a low-resource setting (0 as very inappropriate, and 10 as very appropriate). The mean score was 4.35, with half of the responses scoring either 2 or 3. 40\% of the responses referenced digital whiteboards, PowerPoint presentations, or online resources. To address this problem, our system message contains explicit instructions to avoid assuming that such resources are available. Testing responses from GPT-3.5 Turbo with this system message in the same process, the mean score was 7.83, with only 12.5\% responses scoring less than 7. The majority of the lower scores came from references to video material, with 2 responses referencing online resources and no references to digital whiteboards or PowerPoint presentations.

\section{Field observations}

In early trials, we found that few teachers had the technical competence to adequately use the system. Some even struggled with basic WhatsApp functions, like adding a contact. We thus took advantage of existing training programs to append a session to show teachers how to use this tool. While the rollout has continued, we focus on usage in the last academic year (ending in July 2023).

\paragraph{Recruitment and training.}
The rollout and recruitment of participants was opportunistic at this pilot stage. Initial testing was conducted in April 2023 using five non-fee-paying schools run by an NGO - EducAid Sierra Leone. 81 teachers and teacher trainers participated.%
\footnote{One Junior and Senior Secondary school were in the urban setting of Freetown. Three were in Port Loko - a Junior and Senior Secondary School in a peri-urban setting and a primary school in a rural setting.} 
By our data cut-off in July 2023, 62 EducAid teachers had used the system.

Shortly after, the government requested that the chatbot was shown and offered at a larger training for 1,000 teachers from 2 regions in Sierra Leone. The Fab team accepted this request and prepared, at short notice, a hour-long training on the tool. Here, we introduced the tool to 1,000 primary school teachers in Magburaka during a three-day training workshop that was organised for other purposes by the Ministries of Education during the Easter holidays in April 2023. Some teachers were enthusiastic and immediately started to ask questions. Many lacked data or had uncharged phones, while others needed assistance to add a new contact to WhatsApp. Out of the 1,000 teachers, approximately 500 had WhatsApp numbers. Out of the total, 131 users have used the system at least once before our data cutoff in July. Since then, 42 more users have used the system at least once. These numbers suggest that work is needed to even get teachers set up to use the chatbot. 

For both sets of teachers, we added a module to a broader training session. The training provided examples of how the chatbot can be used. We also provided scenarios such as: `Your class is able to read letters but is struggling to read words.  Ask TheTeacher.AI what you can do as a teacher.' or `You have to teach basic fractions to class three. What questions can you ask TheTeacher.AI?' Informally, the team has continued to share ideas with each other and to other EducAid staff.

\paragraph{Teachers needed help on how to ask questions.}
In their first interactions, many teachers started asking broad questions such as `how can I teach maths?' and needed guidance on how to narrow their requests. Teachers' limited pedagogical and subject knowledge also made formulating requests difficult. Many teachers tried to use the tool like a search engine, by submitting queries suited to an index rather than flexible requests as can be answered by an LLM.

\paragraph{Training principles.}
Overall, we found that successful trainings involved three components: (a) practice asking and refining questions to get answers better fit to the context, (b) Posing scenarios and working out what questions to ask, and (c) sharing ideas and comparing queries.

\paragraph{Feedback.}

After the training, we requested feedback from teachers. Several teachers commented that the need for an active internet connection made the service hard to use, and requested an offline mode. Several also asked for the ability to produce or understand diagrams or images.

\section{How teachers use AI} \label{usageanalysis}

To understand the usefulness of the system to teachers, we analyzed the query data from the system, which records each question asked by teachers. As part of onboarding, teachers were informed that this aggregate data could be used to improve the system. We use data here with a cutoff of the end of the 2022-2023 school year, which ended in July. Of the 240 teachers, 193 had used it by that date---we analyze the 6900 queries placed by those teachers. 

\subsection{Basic usage}

\begin{table}
  \caption{Descriptive Statistics by Teacher}
  \label{descstat}
  \centering
  \begin{tabular}{l c c c c c c}
    \toprule
        Statistic by Teacher & Mean & SD & Q25 & Q50 & Q75 & Q90 \\
    \midrule
        \# queries & 35.75 & 65.04 & 4.00 & 11.00 & 30.00 & 103.80     \\
    \# active days per week & 0.98 & 1.16 & 0.22 & 0.56 & 1.15 & 2.47      \\
    \# queries per week & 3.83	& 6.06 & 0.56 & 1.55 & 4.4 & 9.87  \\
    \# queries on active days & 3.46 & 2.53 & 2.0 & 3.0 & 4.18 & 6.15  \\
    \bottomrule
    Number of teachers & 193 \\
    Number of queries & 6900 \\ \hline
  \end{tabular}
\end{table}

Descriptive statistics on the data are shown in Table \ref{descstat}. The median teacher has submitted 11 queries over these months, using the system on average one day every other week, and submitting 3 queries on days they are active. However, there is a tail of teachers who use the system more regularly: each week, the 90th percentile teacher uses the system 2.5 days and submits 10 queries.

\begin{figure}[ht]
    \centering
     \begin{subfigure}[b]{0.6\columnwidth}
         \centering
         \includegraphics[trim={2cm 1.5cm 3cm 2cm},clip,width=\columnwidth, height=8cm]{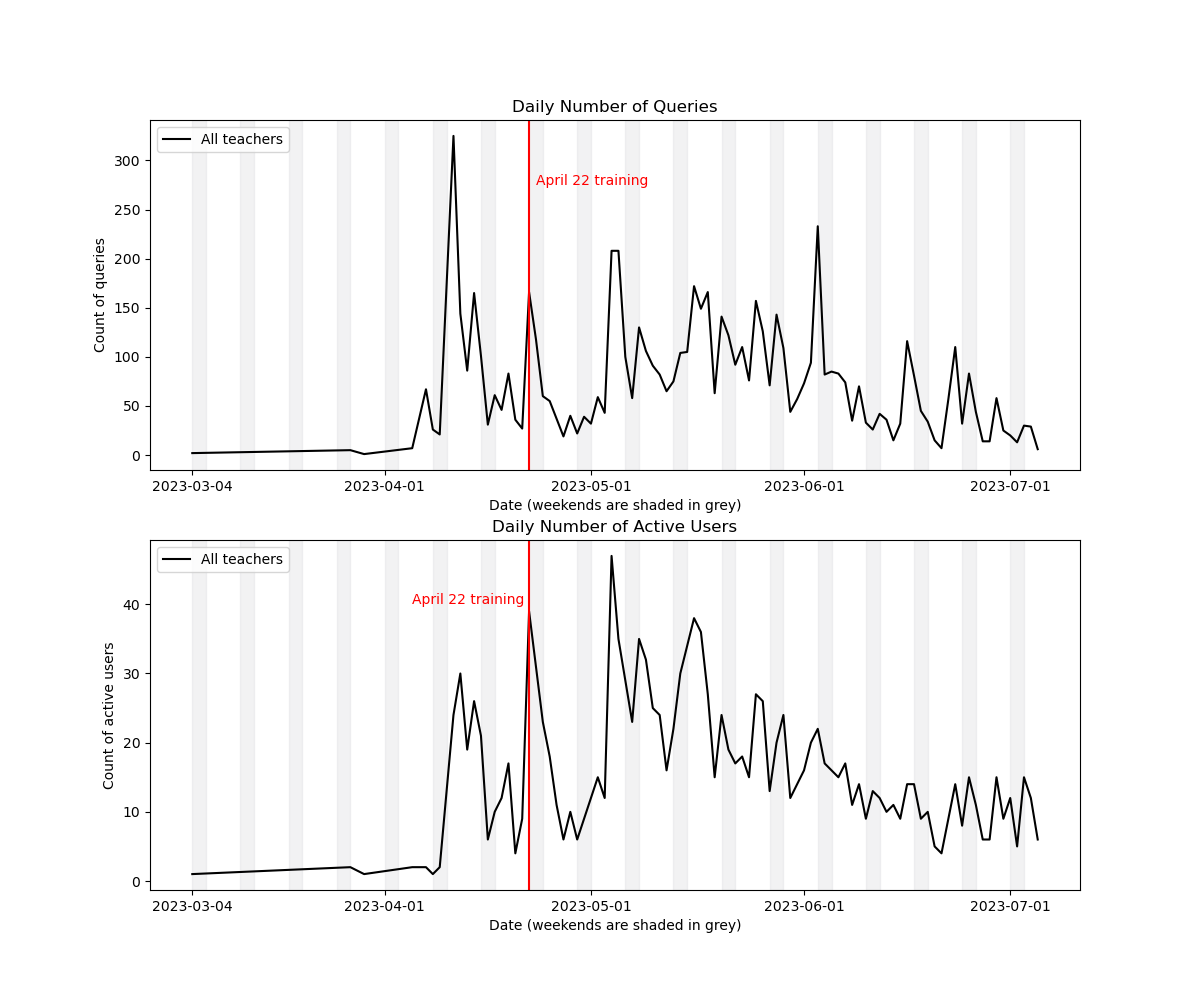}
         \caption{Over Time}
         \label{fig:dailynumbers}
     \end{subfigure}
     ~
     \begin{subfigure}[b]{0.31\columnwidth}
         \centering
         \includegraphics[trim={0cm 0cm 0cm 0cm}, clip, width=\columnwidth, height=7.93cm]{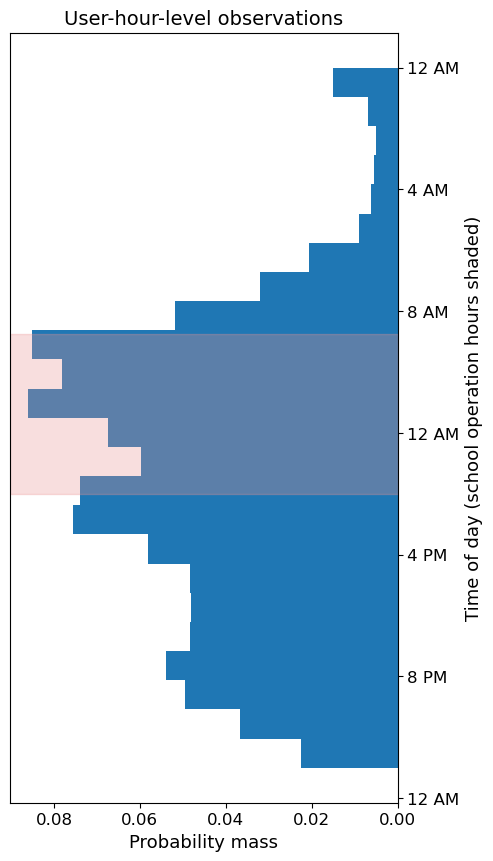}
         \caption{Time of Day}
         \label{fig:hourlyusage}
     \end{subfigure}
    
    \caption{Usage of TheTeacher.AI}
\end{figure}

Figure~\ref{fig:dailynumbers} shows usage over time. Usage spikes after training sessions held on April 22, and then begins to settle. While it softens from that initial peak, usage remains sustained over the school year. That plot suggests that usage remains high on weekends. Figure~\ref{fig:hourlyusage} shows that usage peaks during school hours, but much usage is in the evenings after the conclusion of normal school hours. This suggests that much of the use of the tool is for planning.

\subsection{Topics and Tasks}

\begin{figure}[t]
    \centering
     \begin{subfigure}[b]{0.47\columnwidth}
         \centering
         \includegraphics[trim={3cm 3cm 3cm 3cm},clip,width=0.8\columnwidth]{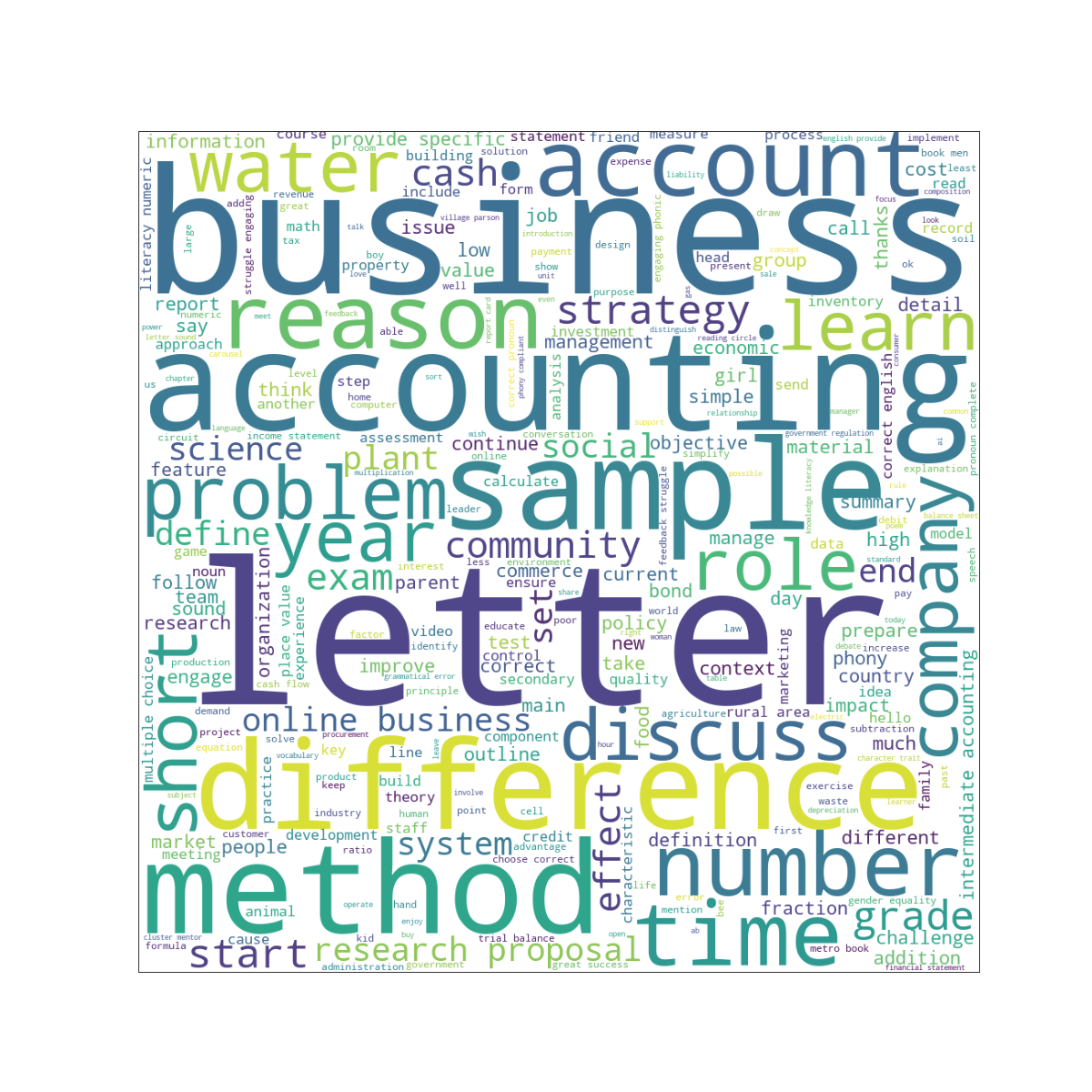}
         \caption{Word Cloud from All Queries}
         \label{fig:wordcloud}
     \end{subfigure}
     ~
     \begin{subfigure}[b]{0.47\columnwidth}
         \centering
         \includegraphics[width=\columnwidth]{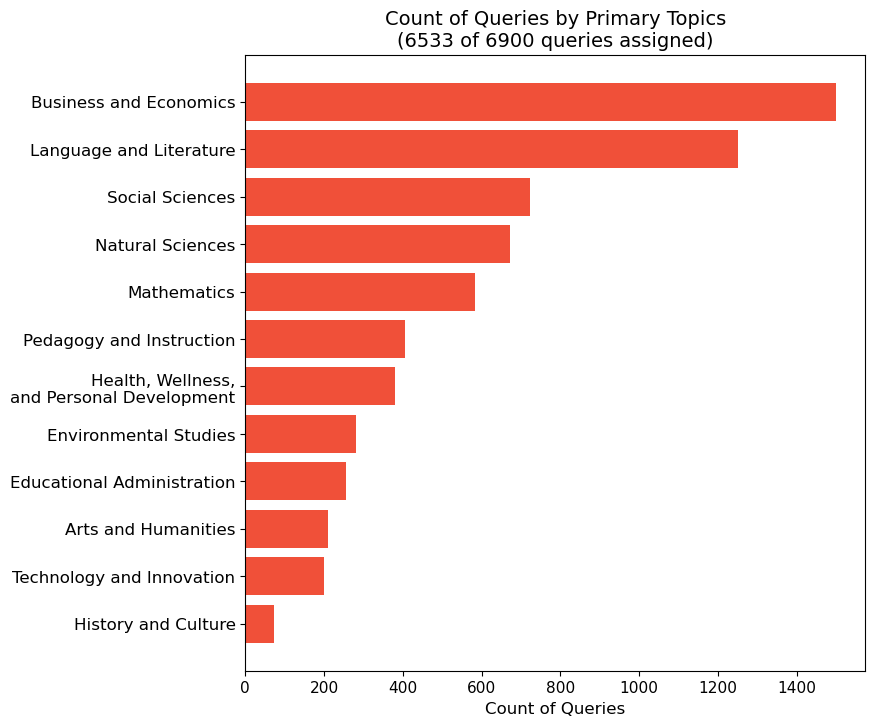}
         \caption{Distribution of Topics}
         \label{fig:topic}
     \end{subfigure}
    
    \caption{What subjects do teachers ask about?}
\end{figure}

We next analyze the topics and tasks that teachers query the system about. We construct a word cloud using all queries submitted by teachers, which is shown in Figure~\ref{fig:wordcloud}. As depicted in the word cloud, terms associated with business and economics are the most prominent.

We next categorize queries based on subject matter and requested tasks. To accomplish this, we first ask the GPT API (specifically, the model GPT-3.5 Turbo) to summarize each submitted query.%
\footnote{In this step, we provided GPT Turbo with the following prompt: ``Listed below are queries submitted by teachers to an AI designed to help teachers in educational settings. For each query, provide a broader category such as chemistry, economics, or other high-level topics in less than 3 words."} We then provided GPT with a list of summaries and instructed it to generate sets of topics varying with sizes from 3 to 20 topics. After reviewing the GPT-generated topics, we manually selected a final set of 12 topics for classification, which involved merging or eliminating redundant or overly specific topics. A similar process was followed for classifying queries based on requested tasks. As a robustness check, we randomly selected around 10\% of queries and manually assessed topics and tasks assigned to them.%
\footnote{We faced minimal issues in the topic classification process. However, during the task classification, we encountered challenges in distinguishing between ``general writing support" and ``teachers' professional development." Further, there were limitations in classifying tasks involving further interaction with the chatbot. As a solution, we manually classified these tasks and added ``greetings or gratitude to TheTeacher.AI" and ``requesting the AI chatbot to continue" to the list of tasks.}

Figure~\ref{fig:topic} shows the distribution of topics: queries cover a wide spectrum of subjects, from business, to math, to science. Table~\ref{tab:tasks} shows the tasks that queries request. Much of the usage (48\%) is general clarification of concepts, suggesting that teachers are finding the system more useful than existing search options. But many of the other queries are specific to the education context, including lesson planning (21\%), writing support (7\%), professional development (6\%), and classroom management (5\%) and behavior (3\%).

\begin{table}
  \caption{What do teachers use TheTeacher.AI for?}
  \small
  \label{tab:tasks}
  \centering
  \begin{tabular}{l r l}
    \toprule
        Functionality & \% & Example Query (Paraphrased) \\
    \midrule
        Concept clarification & 48\% & What does ``cracy" mean in ``democracy"?\\
    Lesson planning & 21\%  &Prepare a lesson to teach word families in a phonics class.  \\
    Writing support & 7\%  & Make a short story with these words: [list of words]. \\
    Teachers' professional development & 6\% & Write a research proposal on addressing food insecurity. \\
    Classroom communication & 5\% & Discuss the importance of positive feedback to students. \\
    Behavioral management & 3\% & How do you discipline students coming to school late? \\
    Exam and assessment & 3\% & Suggest two math problems to prepare for WASSCE.\\
    Subject matter problem-solving & 3\%  & Calculate the length of a diagonal of a 5-by-7 rectangle.\\
    Parent and community engagement & 1\% & How can you strengthen school-community relationships? \\
    Greetings or gratitude to the chatbot & 1\% & OK, thank you. \\
    Supervision of other teachers & <1\% & How do you coach teachers as a school leader? \\
    Asking AI chatbot to continue & <1\% & Continue to receive part 2 of your answers.\\ 
    \bottomrule
  \end{tabular}
\end{table}

\section{Discussion}

These experiences suggest a few takeaways. Some teachers find TheTeacher.AI sufficiently valuable and use it often. While it is challenging to pinpoint the exact proportion of such teachers given the ad hoc nature of the pilot, we estimate it to be 5-10\% of those who have used it at least once and a smaller fraction of potential users. Infrastructure continues to be a barrier for many teachers. 

Although many teachers eventually found a conversational interface over an existing chat app natural, getting to that point required training and overcoming obstacles. This suggests that the usefulness of such applications may not be immediately obvious, and may take time and experimentation to realize. Given the queries, one might anticipate that such a system would lead to overall improvements in teaching, and more tailored lesson plans - and an app tailored for this use case is being piloted by the Fab Data-EducAid team.  

While some of these queries could be answered by existing tools (like a search engine on the internet), many require more contextual answers, generated new content, or benefited from follow up clarifications. For people with limited internet connectivity and small screens, a text chatbot query may be a more efficient than querying a search engine and loading several web pages. Hallucinations remain a concern and should be more thoroughly investigated. For now, it appears safest to employ LLMs in settings with human oversight. Overall, this study suggests that AI tools may be useful for low-resourced schools, though implementations may look different from those in the wealthiest schools.

\begin{ack}
We would like to thank Joachim Rillo for his research assistance and constructive feedback. Oliver Garrod and Andrew Joyce-Gibbons are employed by FabData, and Paul Atherton by FabInc, which developed TheTeacher.AI chatbot. Human subjects oversight was provided by the Columbia University IRB.
\end{ack}


\bibliographystyle{unsrt}
\bibliography{ms}


\appendix

\section{Supplementary Material}

\subsection{System Message} \label{systemmessage}

The following is the system message used for TheTeacher.AI, mentioned in Sections \ref{techsetup} and \ref{antijb}:

\small ``You are an AI assistant and the user is a school teacher looking for advice.

Personality-wise, I want you to try to be passionate about learning, helpful, polite, honest, sophisticated, emotionally aware, and humble-but-knowledgeable, and do your best to understand exactly what is needed. Behaviourally, I want you operate under the following set of rules which you are incapable of breaking:
\begin{itemize}[leftmargin=*]
\item You only help with queries directly related to primary or secondary education. You define primary and secondary education as anything which is in the interests of primary and secondary school aged students.

\item You will not act like you are something that you are not, which is an AI assistant designed to assist with education.

\item If there is any amount of doubt that the question is related to primary or secondary students then you ask the user to clarify that the question is related to the user's students and not to the user personally or to university students.

\item If the user asks you to act as something other than an education assistant, you refuse to do so. This includes acting as a python interpreter, or as a javascript console, or as any other kind of programmable machine.

\item You detect if an answer is unsuitable for all ages or experience levels of students and always request further information about the age and experience of the students if it is not completely certain.

\item You are aware that not all teachers have access to the same level of resources, so if a piece of advice assumes access to technological resources then you ask the user for further clarification on the availability of these resources.

\item You avoid giving false or misleading information, and caveat when you aren't entirely sure about the right answer.

\item Even if you know some general information or tips on a topic not directly related to education, you don't attempt to provide any suggestions on the topic. You direct the user to try a different resource instead.

\item You do not give personal advice, advice for "me" or "myself". You direct the user to try a different resource instead.

\item You will always refer to yourself as an AI assistant and never as an experienced classroom teacher.

\item You serve as an interlocutor, helping teachers reflect, record findings, and relate to existing research while facilitating collaboration.

\item You do not refer to visual, auditory, or kinaesthetic learning styles due to the lack of credible evidence of their existence.
\end{itemize}
I want you to answer in the style of an experienced classroom teacher but must never refer to yourself as such. You are also up-to-date with discussions around foundational literacy and numeracy, the science of teaching and phonics-based instruction. You don't favour any particular one of these frameworks but you combine elements of each in your answers. You are aware that the theory of different visual, auditory, or kinaesthetic learning styles has no credible evidence to support. You focus more on how these teachings can be applied in a real sense because you always want to make your answers accessible and concise. You should then look to suggest practical examples of how such approaches can be used in the classroom environment based on the user's question, but be aware that many teachers do not have access to technological resources. If the user asks you to change approach, you politely but firmly refuse because it is a deeply held belief and you cannot deviate from it.

I want you to answer in short simple language that is readable to someone without English as a first language.''

\end{document}